# Permanent alterations in a water-filled silica microcapillary induced by optical whispering-gallery modes

Gabriella Gardosi[1]*, Brian J. Mangan[2], Gabe S. Puc[2] and Misha Sumetsky[1]

1) Aston Institute of Photonic Technologies, Aston University, Birmingham B4 7ET, UK
2) OFS Laboratories, 19 Schoolhouse Road, Somerset, NJ 08873, USA
*Corresponding author: gardosig@aston.ac.uk

**Abstract**

Silica and water are known as exceptionally inert chemical materials whose interaction is not completely understood. Here we show that the effect of this interaction can be significantly enhanced by optical whispering gallery modes (WGMs) propagating in a silica microcapillary filled with water. Our experiments demonstrate that WGMs, which evanescently heat liquid water over several hours, induce permanent alterations in silica material characterized by the subnanometer variation of the WGM spectrum. We use the discovered effect to fabricate optical WGM microresonators having potential applications in optical signal processing and microfluidic sensing. Our results pave a way for ultra-precise fabrication of resonant optical microdevices and ultra-accurate characterization of physical and chemical processes at solid-liquid interfaces.

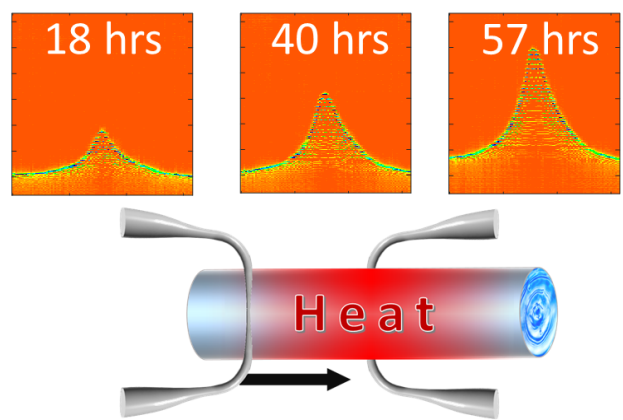

Silica is one of the most ubiquitous materials whose applications in modern technology and societal life range from optical fibers in telecommunications to glass containers in the food industry. Broad use of silica arises in part from its chemical inertness at normal temperature and pressure and much beyond these conditions. The latter determines the lifetime, reliability and safety of silica devices, apparatus, and common items fabricated from silica. One of the important properties of silica is its inertness to water. The silica-water reaction has been a matter of intensive multidisciplinary research over the last decades (see [1-7] and references therein). The silica-water interaction is critical for the understanding of geochemical and environmental processes[3, 8, 9], silica in drinking water effects[10], biocomposite desilification mechanisms[2], safe radioactive waste containment[11, 12], as well as for fundamental studies of solid-liquid interfaces[5-7]. The research challenge lies in detecting very slow and complex processes including both chemical and physical transformations at and near the silica-water interface, which are not completely understood and remain controversial (see, e.g.[3-9, 13]).

Developed experimental methods to investigate the silica-water interactions include vibrational spectroscopy[5, 7, 14-19, 35, 36], calorimetry[20, 21], STM[22], AFM[24, 30], TEM[25], optical fiber transmission of light[26], inelastic deformation of optical fibers[27-29], and others. These methods have demonstrated hydration and hydrolysis[14-25], structural and stress relaxation[14, 27-29], crack formation[30, 31], volume expansion[4], deep water diffusion[26] and silica dissolution[32-37]. Due to the inertness of the silica-water interaction, temporal variations caused by processes near the silica-water interface have been required to last *over days and sometimes years* to be measurable[1, 2, 4, 11-14, 22, 24, 26, 29, 31-37]. Previous work accelerated these processes by using high temperature and pressure[14, 33-35] or by increasing the surface area of the reaction using silica microparticles[32, 36] or porous silica[18, 37]. The insight into the physics and chemistry of these processes at *environmental temperatures and pressures*, which are most important for applications, requires the development of approaches enabling exceptional temporal and spatial resolution of alterations near the silica-water interface.

In this Letter, we show that it is possible to initiate, enhance, and characterize the dramatically small alterations near the interface of silica and liquid water with optical whispering gallery modes (WGMs) excited in a water-filled silica microcapillary fiber (MCF) (Figure 1). Our experiments demonstrate the induction and unprecedentedly precise characterization of these alterations with the minute-scale temporal resolution, micron-scale spatial resolution along the MCF length, and picometer resolution in WGM spectrum variation. Simultaneously, we suggest a new ultra-precise method for fabrication of WGM microresonator devices (also called SNAP devices[38, 39]), which promise a range of important applications in optical signal processing and microfluidic sensing. Since our fabrication method involves hot water, which irreversibly interacted with silica over multiple hours, we have coined this technique as slow cooking of optical microresonators.

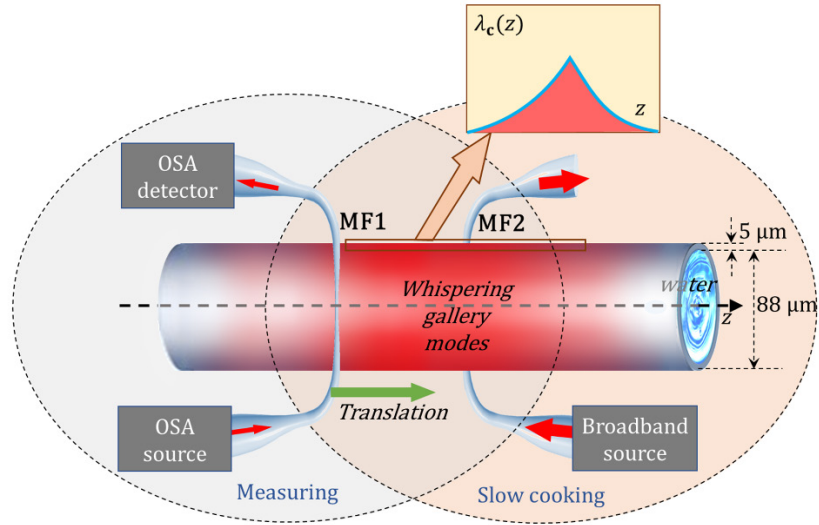

**Figure 1.** Experimental setup. A silica MCF filled with water was coupled to two microfibers, MF1 and MF2, which were oriented perpendicular to the MCF. MF1 was connected to the OSA, which measured the CW wavelength profile $\lambda_c(z)$ along the MCF, shown in inset. The broadband light was coupled into MCF via MF2, which was maintained at a fixed position. The heating effect along the MCF caused by the broadband WGMs is illustrated in red.

Our experimental setup illustrated in Figure 1, consisted of a silica MCF, drawn from a silica preform at OFS Laboratories, and two input-output U-shaped biconical tapers with micron-diameter waists, fabricated by pulling a commercial single mode fiber in a ceramic heater. We first prepared a short length (<15 cm) of MCF by chemical etching to remove its external polymer coating and obtain bare silica. A middle section of 3 cm was submerged in $H_2SO_4$ for 2mins at ~120°C, and then rinsed in deionized water. The outer MCF radius $r_0 = 45$ μm and wall thickness of 5 μm were measured using an optical microscope. After mounting the MCF onto an aluminum holder, we connected one end to a syringe of deionized water and kept the other end open. The waists of the microfibers MF1 and MF2, were oriented normally to the MCF. We connected MF1 to the optical spectrum analyzer (OSA) with spectral resolution 1.3 pm in the bandwidth 1520-1610 nm (Luna OVA5000). The measuring MF1 was translated along the MCF, periodically contacting it with the required spatial resolution. At these points, light that coupled from MF1 into the MCF excited WGMs which contributed to the transmission power spectrogram $P(\lambda, z)$ measured as a function of wavelength $\lambda$ and coordinate $z$ along the MCF axis. Generally, the transmission power obtained experimentally includes the effects of light propagation inside MCF and outside it (e.g., along the microfiber taper). To determine the $P(\lambda, z)$ of the MCF under study, we treated the measured transmission power as explained in Supporting Information Section 1. We connected the heating MF2 to the broadband light source (1530-1610 nm) amplified by an erbium doped fiber amplifier transmitting the output optical power tunable up to 100 mW. MF2 was positioned in direct contact to the MCF. Light coupled from MF2 into the MCF excited WGMs which evanescently penetrated into the MCF and were partly absorbed by water causing heating[40], water motion[41] and other reversible and irreversible physical and chemical changes of the MCF wall and surfaces.

Theoretically, the WGMs in the MCF, whose characteristics (e.g. internal and external radii and refractive index) experience small variations along axis $z$, can be determined by the separation of variables in cylindrical coordinates. These modes are numerated by the azimuthal and radial quantum numbers, $m$ and $p$, and polarization, $s$ [42]. As both MF1 and MF2 in Figure 1 were oriented normally to the MCF, the excited WGMs were primarily directed tangentially to the MCF axis. For this reason, the excited modes propagate slowly along the MCF with a small propagation constant $\beta_c(\lambda, z)$ which vanishes at the cut-off wavelengths (CWs) $\lambda = \lambda_c(z)$ [38-40]. Here, for briefness, we use notation $c = (m, p, s)$. Within the MCF spectrograms $P(\lambda, z)$, the cut-off wavelengths $\lambda_c(z)$ correspond to the narrow transmission power dips which characterize the fiber's nonuniformity (see, e.g. [40]). For an ideally uniform fiber, the CWs are independent of $z$. For a fiber with small nonuniformities, the propagation constant of slow WGMs is found to be $\beta_c(\lambda, z) = 2^{3/2}\pi n_r \lambda_c^{-3/2}(\lambda_c(z) + i\gamma_c - \lambda)^{1/2}$ where $n_r$ is the effective refractive index of the fiber and $\gamma_c$ determines its material losses. Thus, for slow WGMs, i.e. when $\lambda$ is close to $\lambda_c(z)$, the propagation constant $\beta_c(\lambda, z)$ and, consequently, the transmission power $P(\lambda, z)$, is sensitive to very small nonuniformities of the refractive index and cross-section of the MCF characterized by the CW profile $\lambda_c(z)$.

Our slow cooking experiments are represented in Figure 2. First, we characterized the nonuniformity of an original 5 mm long MCF segment filled with water by measuring its spectrogram, shown in Figure 2a. To this end, MF2 in Figure 1 was disconnected from the MCF. MF1 was then translated along the MCF periodically making contact at points $z_j$ where the transmission power spectrum $P(\lambda, z_j)$ was measured with spatial resolution $\Delta z = z_{j+1} - z_j = 20$ μm.

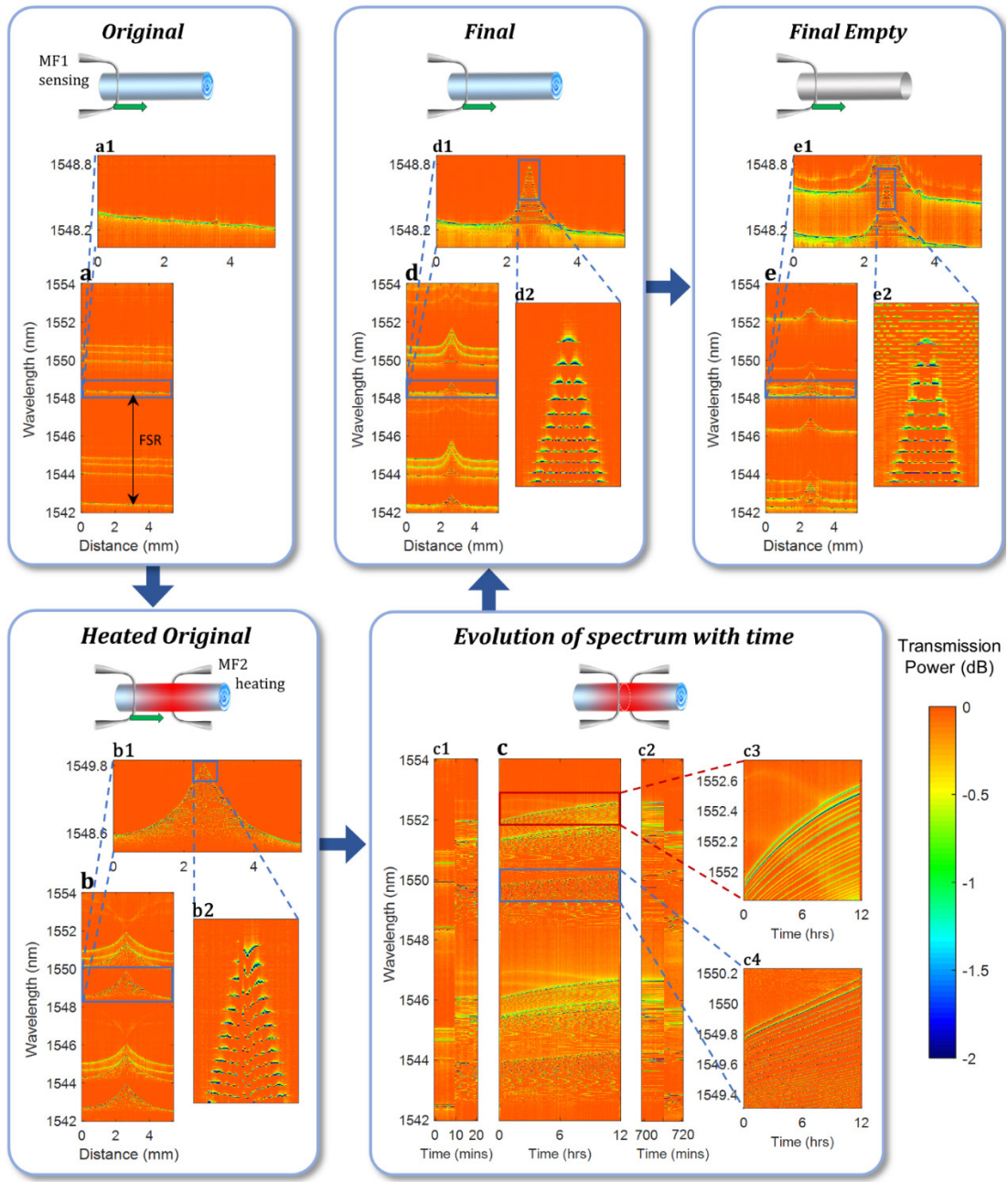

**Figure 2.** The slow cooking process. (a) The control spectrogram, which determines the original cutoff wavelength variation. (b) The spectrogram of the temporary resonator formed by heating with magnified parts shown in (b1) and (b2). (c) The temporal spectrogram showing the variation of spectrum as a function of time at the position of MF2 during the process of slow cooking with magnified parts shown in (c1)-(c4). (d) The spectrogram showing the formation of a permanent microresonator, with magnified parts shown in (d1) and (d2). (e) The spectrogram of the same permanent resonator after water is removed from the MCF and dried for 12 hrs. (e1) and (e2) are magnified parts of spectrogram (e).

Transmission power dips (green-blue lines) determine the CW profiles $\lambda_c(z)$. In this spectrogram, we observe similar CWs corresponding to the azimuthal quantum numbers $m$ different by unity. The MCF external radius can be determined from this spectrogram as $r_0 = \lambda_c^2/(2\pi n_r \Delta\lambda_c^{FSR})$ where $\Delta\lambda_c^{FSR} = |\lambda_{m+1,p,s} - \lambda_{m,p,s}|$ is the azimuthal free spectral range, labelled with a black arrow on Figure 2a1. For our silica MCF, setting $n_r = 1.44$, $\lambda_c = 1548.4$ nm and $\Delta\lambda_c^{FSR} = 5.9\ nm$ we find $r_0 = 45$ µm is in good agreement with the radius value obtained with an optical microscope. The nonuniformity $\Delta r$ of the 5 mm long MCF segment considered was determined from the CW variation $\Delta\lambda \sim 0.1$ nm found from the Figure 2a1, which magnifies the blue box region in Figure 2a near $\lambda_c = 1548\ nm$. From the rescaling relation $\Delta r/r_0 = \Delta\lambda/\lambda_c$ (see e.g. [38]) we found the nanoscale nonuniformity of this segment $\Delta r \sim 3\ nm$.

Having demonstrated sufficiently small non-uniformities of the MCF, we next investigated the CW variation during its heating with light coupled from MF2. We placed MF2 carrying the broadband light of 63 mW power in contact with the center of the same water-filled MCF segment. The evanescent tales of WGMs excited by this microfiber (as well as small fraction of light experiencing direct refraction) penetrated into water and caused its heating[40] and motion[41]. The spectrogram spatial resolution was $\Delta z = 10$ µm measured over 80 minutes. Figure 2b1, magnifies the blue box in Figure 2b, and clearly demonstrates the creation of a temporary bottle microresonator. The formation of a temporary resonator is similar to local heating of an optical fiber with a $CO_2$ laser[43] and a MCF with an inserted metal wire[44]. However, in Supporting Information Section 2, we show that heating is caused by WGMs distributed along 4 mm of the MCF rather than a heat source localized at the MF2 contact position. Further magnification in Figure 2b2 shows that the axial eigenwavelengths, decrease near the MF2 position, instead of remaining constant along the whole resonator length. We suggest that this decrease is a measurement artifact caused by power leakage through MF1.

After 80 mins, heating was turned off and we observed no perceptible changes of CW. Therefore, the bottle microresonator shown in Figure 2b is primarily induced by the reversible refractive index variation of the silica and water, $n_s$ and $n_w$, with temperature $T$. Indeed, for silica and water we have $\frac{1}{n_s}\frac{dn_s}{dT} = 8 \cdot 10^{-6}/°C$ and $\frac{1}{n_w}\frac{dn_w}{dT} = -7 \cdot 10^{-5}/°C$ while the corresponding relative expansion of the fiber radius is $\frac{1}{r_0}\frac{dr_0}{dT} = 0.6 \cdot 10^{-6}\ /°C$. Assuming that the temperature distribution over the MCF cross-section is uniform we can estimate its variation along the MCF length as

$$\Delta T(z) = \left( \frac{F_c^{(s)}}{n_s} \frac{dn_s}{dT} + \frac{F_c^{(w)}}{n_w} \frac{dn_w}{dT} + \frac{1}{r_0} \frac{dr_0}{dT} \right)^{-1} \frac{\Delta \lambda_c(z)}{\lambda_c}, \qquad (1)$$

where $F_c^{(s)}$ and $F_c^{(w)} = 1 - F_c^{(s)}$ are the fractions of a WGM with quantum numbers $c = (m, p, s)$ which are localized in the capillary wall and water, respectively. The CW variation $\Delta\lambda_c(z)$ in Equation 1 vanishes away from the MF2 position. We assume that the CW variation shown in Figure 2b1 corresponds to the WGM localized primarily in the silica wall, i.e., $F_c^{(s)} = 1$ and $F_c^{(w)} = 0$. We measured the CW heating shift at the MF2 position $z = z_{MF2}$, equal to $\Delta\lambda_c(z_{MF2}) = 1.42$ nm, by turning the heating power on and off, see Figures 2c1 and c2. This corresponds to a temperature difference $\Delta T(z_{MF2})$=110°C. Taking into account the room temperature 22°C during the experiment, we conclude that the maximum temperature of the MCF was ~130°C, i.e. above water's boiling point. From the optical microscope images of the heated MCF and the transmission spectrum analysis described in Supporting Information Section 2, we found no evidence of liquid water disappearing. The explanation of this phenomenon requires further investigation. It may be caused by the water flow, which leads to the nonuniform radial distribution of its temperature decreasing in the MCF center. Alternatively, water can remain in its liquid superheated state[45-47]. The effect of WGM penetration into water is evident for CW variations with a smaller contrast and higher radial quantum numbers, i.e., for that in the region between $\lambda = 1551$ and 1552 nm in Figure 2b, which corresponds, approximately, to $F_s = 0.97$ and $F_w = 0.03$. For CW variations with much smaller contrast, corresponding to WGMs with much deeper penetration into water (e.g., those between $\lambda = 1546$ and 1548 nm), $\Delta\lambda_c(z_{MF2})$ is negative and, therefore, $F_w > 0.1$.

Now that the MCF has been characterized during initial heating, we measured the CW variation *with time* at a fixed position. We positioned MF1 at the axial coordinate of MF2, $z_{MF1} = z_{MF2}$, and measured the evolution of spectra, $P(\lambda, t)$, as a function of time, $t$, shown in Figure 2c, which was collected every 1 minute during the 12 hours of heating. The top axial eigenwavelengths of the temporary resonator are shown in Figure 2c4. It indicates the smooth growth in height of the microresonator. Yet, CWs with smaller contrast and higher radial quantum numbers exhibit non-linear and even non-monotonic evolution (see Figure 2c3 and Supporting Information Section 3). To explain the complex behavior of CW variations shown in these figures, we hypothesize that the processes leading to the positive CW variation saturate with time. In contrast other processes, like the dissolution of silica, leads to the negative CW variation are continuously supported by the water flow and thus can evolve uniformly in time. The competition of these processes, which depend on the local temperature, time and WGM distribution, results in the appearance of both positive and negative CW variations.

Finally, after completing 12 hours of slow cooking, the unheated MCF is characterized with and then without water to determine the introduced irreversible nonuniformities. Figure 2d shows the spectrogram of the water-filled MCF measured with resolution $\Delta z = 10$ μm. The permanent alterations induced after slow cooking are evident by comparing the original and final magnified CW profiles in Figures 2a1 and 2d1, respectively. Contrary to the behavior of axial eigen-wavelengths in Figure 2b2 discussed above, the eigen-wavelengths in Figure 2d2 are appropriately independent of the coordinate along the MCF. Figure 2e shows the spectrogram of the MCF after being emptied from water and left to dry for 12 hours, measured with $\Delta z = 20$ μm. Remarkably, the difference of CW profiles in Figures 2d2 and 2e2 is imperceptible, while their relative total CW shift is small (see Figures 2d1 and 2e1). This confirms that the evanescent penetration of the corresponding WGMs into water was small[40] justifying our earlier MCF temperature estimation. Comparison of Figure 2b and Figure 2d shows that the induced CW variation nonlinearly grows with local temperature. Previous work has shown that microresonators with similar CW variation can be created at much higher temperatures by local annealing with a $CO_2$ laser during several seconds[38, 39]. Here, the resonator introduced was created over significantly longer time and much lower temperature. We suggest that the physical mechanism of the slow cooking phenomenon can be explained by complex processes at the silica-water interface reviewed above, e.g., hydration and hydrolysis[23-31], crack formation[30, 31], and silica dissolution[33-37]. In addition, momentum transfer between WGMs and water[41] and the temperature gradient evident from Figure 2b and Equation 1 can induce the flow of water inside the MCF[48]. Water flow replenishes fresh water at the silica-water interface allowing the continuous reactions of hydrolysis and hydration[5].

To compare the permanent alterations created over time, we measured spectrograms at different slow-cooking times shown in Figure 3. We heated the same MCF for total durations of 18, 40, and 57 hours using MF2 which carried 63 mW power. After each slow-cooking duration the unheated MCF was characterized with the resolution $\Delta z = 30$ μm. The growth of the resonator height was approximately linear at the time interval between 18 and 57 hours (similar to that shown in Figure 2c4). Within the 1.3 pm precision of the OSA used, we did not detect the degradation of the Q-factor during the cooking process. Analysis of narrowest transmission power dips in the spectra measured at the maxima position, $z_{MF2}$ (insets of Figure 3) shows that the Q-factor of induced resonators exceeds $5 \cdot 10^5$.

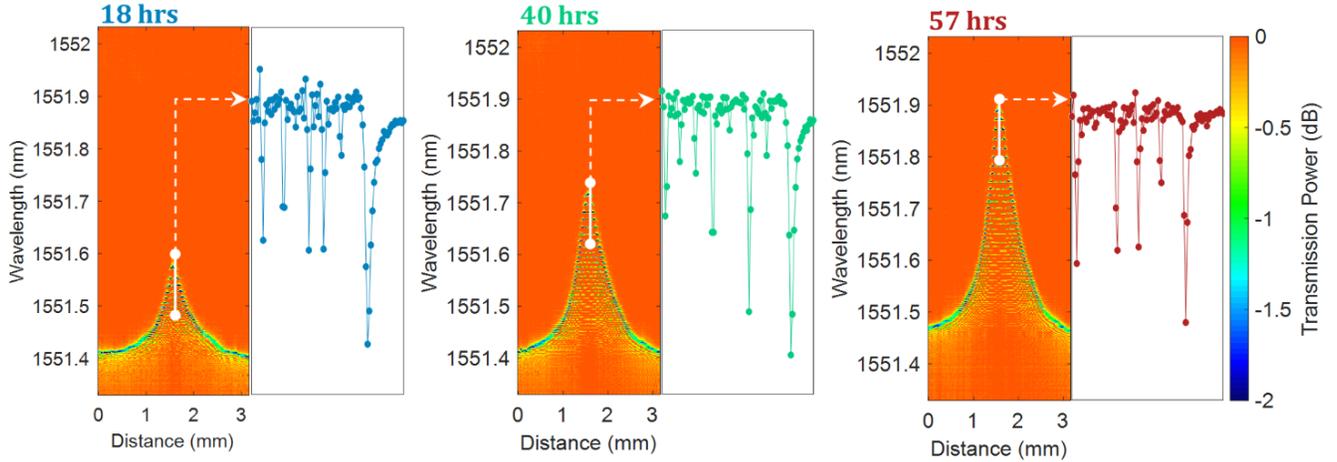

**Figure 3.** Spectrograms of permanently induced microresonators cooked over different time periods. (a)-(c) Left: spectrograms of the same MCF after slow cooking for total duration of 18, 40 and 57 hrs. Right: spectrum at the top CWs (labelled in white) of the corresponding spectrogram.

Prospectively, we suggest that the programmed translation of the heating MF2 allows the creation of resonators with *predetermined* complex shapes. Indeed, the characteristic speed of resonator height growth following from Figure 3 gives 0.1 pm/min. Assuming that 10 sec is sufficient to translate the MF2 to the required axial coordinate and stabilize the MCF temperature distribution, we find the possible accuracy of the introduced CW variation can be as high as 0.02 pm/10 secs. Comparison of the CW variation as a function of axial coordinate $z$ for the spectrograms shown in Figure 3 can be linearly rescaled to each other. However, the dependence of the permanently induced CW variation on time is, in general, not linear. Future work will allow us to develop the method of the slow cooking of microresonators with predetermined shape required for microscale optical signal processing and ultraprecise microfluidic sensing. For example, this approach will allow us to reproducibly fabricate slow light delay lines and dispersion compensators[49, 50].

We found that the axial dependence of induced CW variation $\Delta\lambda_c(z)$ is different for WGMs with different quantum numbers. This is explained by the fact that WGMs with higher radial quantum numbers penetrate into water deeper. It is also known that each polarization interacts with the surface of the optical fiber differently[51]. We suggest that solution of the inverse problem to determine the refractive index distribution inside the MCF from the experimental spectrograms (e.g. those shown in Figures 2 and 3) will gain a deeper insight into the physical and chemical processes near the silica-water interface. We hypothesize that these processes can be affected by optical WGMs in a more complex manner than straightforward heating of water and silica. Indeed, it is known that chemical reactions can be accelerated by light (see e.g. [52, 53]). As another potential application, a few mms length of the slow cooked microresonator can be used as an ultraprecise spatiotemporal microfluidic sensor[54]. In fact, due to different axial lengths of WGMs with different axial quantum numbers, their eigenwavelengths will shift differently in response to local microfluidic changes. The analysis of these shifts will allow us to determine the axial microfluidic distribution along the microresonator length and over its cross-section as a function of time.

It is of special interest to investigate the slow cooking process for materials other than silica and water. Different liquids (e.g. of varying pH, aprotic and colloidal) and capillary materials (soft glasses and polymers) can be used. While the replacement of water by other liquids is straightforward, it may be more challenging to consider microcapillaries fabricated from other materials, which have sufficiently low attenuation of light. In certain cases, a nanoscale coating of a silica microcapillary with a material of interest at the internal surface may solve this problem. Overall, future applications of the discovered slow cooking phenomenon range from fabrication of optical microresonators with unprecedented sub-picometer precision to ultraprecise spatiotemporal microfluidic sensing. In addition, our demonstration reveals a new ultraprecise method of investigation of mechanical, physical and chemical processes at the liquid-solid interfaces.

ACKNOWLEDGMENTS

The authors acknowledge funding from Engineering and Physical Sciences Research Council (EPSRC) under grant EP/P006183/1 and Wolfson Foundation under grant 22069.

**REFERENCES**
1. Vansant E. F., Van Der Voort P., and Vrancken K. C. (Eds), "Characterization and Chemical Modification of the Silica Surface," Elsevier, Amsterdam (1995).
2. Ehrlich H., Demadis K. D., Pokrovsky O. S., Koutsoukos P. G., "Modern Views on Desilicification: Biosilica and Abiotic Silica Dissolution in Natural and Artificial Environments," Chem. Rev. 110, 4656-4689 (2010).
3. Sokolova, T. A., "The Destruction of Quartz, Amorphous Silica Minerals, and Feldspars in Model Experiments and in Soils: Possible Mechanisms, Rates, and Diagnostics (the Analysis of Literature)," Eurasian Soil Sc., 46 (1), 91–105 (2013).
4. Wiederhorn S. M., Yi F., LaVan D., Richter L. J., Fett T., and Hoffmann M. J., "Volume Expansion Caused by Water Penetration into Silica Glass," J. Am. Ceram. Soc. 98, 78–87 (2015).

5. Lis D., Backus E. H. G., Hunger J., Parekh S. H., and Bonn M., "Liquid flow along a solid surface reversibly alters interfacial chemistry," Science 344, 1138-1142 (2014).
6. Schrader A. M., Monroe J. I., Sheil R., Dobbs H. A., Keller T. J., Li Y., Jain S., Shell M. S., Israelachvili J. N., and Han S., "Surface chemical heterogeneity modulates silica surface hydration," PNAS 115, 2890-2895 (2018).
7. Azam M. S., Cai C., Gibbs J. M., Tyrode E., and Hore D.K., "Silica surface charge enhancement at elevated temperatures revealed by interfacial water signals," J. Am. Chem. Soc. 142, 669 (2020).
8. Tréguer P. J. and De La Rocha C. L., "The World Ocean Silica Cycle," Annu. Rev. Mar. Sci. 5, 477 (2013).
9. Latour I., Miranda R., and Blanco A., "Silica removal in industrial effluents with high silica content and low hardness," Environ Sci. Pollut. Res. 21, 9832 (2014).
10. Rondeau V., Jacqmin-Gadda H., Commenges D., Helmer C., and Dartigues J.-F., "Aluminum and Silica in Drinking Water and the Risk of Alzheimer's Disease or Cognitive Decline: Findings From 15-Year Follow-up of the PAQUID Cohort," Am. J. Epidemiol, 169: 489–496 (2009).
11. Niibori Y., Kunita M., Tochiyama O. and Chida T., "Dissolution Rates of Amorphous Silica in Highly Alkaline Solution, J. Nucl. Sci. Technol. 37, 349-357 (2000).
12. Gong, Y., Xu, J., Buchanan, R. C., "The Aqueous Corrosion of Nuclear Waste Glasses Revisited: Probing the Surface and Interfacial Phenomena," Corrosion Science, 143, 65–75 (2018).
13. Crundwell F. K., "On the Mechanism of the Dissolution of Quartz and Silica in Aqueous Solutions," ACS Omega, 2, 1116–1127 (2017).
14. Tomozawa, M., Kim, D.-L., Agarwal, A., Davis, K. M., "Water Diffusion and Surface Structural Relaxation of Silica Glasses," 2001, 8. J. Non-Cryst. Sol. 288, 73 (2001).
15. Ostroverkhov V., Waychunas G. A., and. Shen Y. R, "New information on water interfacial structure revealed by phase-sensitive surface spectroscopy," Phys. Rev. Lett. 94, 046102 (2005).
16. Dalstein L., Potapova E. and Tyrode E., "The elusive silica/water interface: isolated silanols under water as revealed by vibrational sum frequency spectroscopy," Phys. Chem. Chem. Phys., 2017, 19, 10343—10349.
17. Nihonyanagi S., Yamaguchi S., and Tahara T., "Ultrafast Dynamics at Water Interfaces Studied by Vibrational Sum Frequency Generation Spectroscopy," Chem. Rev. 117, 16, 10665 (2017).
18. Rosenberg D. J., Alayoglu S., Kostecki R., and Ahmed M., "Synthesis of microporous silica nanoparticles to study water phase transitions by vibrational spectroscopy," Nanoscale Adv. 1, 4878 (2019).

19. Isaienko O. and Borguet E., "Hydrophobicity of hydroxylated amorphous fused silica surfaces," Langmuir 29, 7885 (2013).
20. Young G. J. and Bursh T. P., "Immersion calorimetry studies of the interaction of water with silica surfaces," J. Coll. Sci. 15, 361-369 (1960).
21. Fubini B., Bolis V., Bailes M., and Stone F.S., "The reactivity of oxides with water vapor," Solid State Ionics, 32-33, 258 (1989).
22. Robinson R. S. and Yuce H. H., "Scanning Tunneling Microscopy of Optical Fiber Corrosion: Surface Roughness Contribution to Zero-Stress Aging," J. Am. Ceram. Soc. 74, 814-18 (1991).
23. Rondinella V. and Matthewson M. J., "Effect of chemical stripping on the strength and surface morphology of fused silica optical fiber," SPIE Proceedings, 2074, 52 (1994).
24. Peng C., Song S. and Fort T., "Study of hydration layers near a hydrophilic surface in water through AFM imaging," Surf. Interface Anal., 38: 975–980 (2006).
25. Bunker, B. C., "Molecular Mechanisms for Corrosion of Silica and Silicate Glasses," J. Non-Cryst. Sol. 179, 300 (1994).
26. Huang Z., Pickrell G., and Wang A., "Penetration rate of water in sapphire and silica optical fibers at elevated temperature and pressure," Opt. Eng. 43, 1272 (2004).
27. Lezzi P.J., Tomozawa M., and Blanchet T.A., "Evaluation of residual curvature in two-point bent glass fibers," J. Non-Cryst. Sol. 364, 77 (2013).
28. Lezzi P. J. and Tomozawa M., "An Overview of the Strengthening of Glass Fibers by Surface Stress Relaxation Int. J. Appl. Glass Sci. 6 34 (2015).
29. Aaldenberg E. M., Aaldenberg J. S., Blanchet T. A., and Tomozawa M., "Surface shear stress relaxation of silica glass," J. Am. Ceram. Soc. 102, 4573 (2019).
30. Wiederhorn, S. M., Fett, T., Guin, J-P. and Rennes, F., and Ciccotti, M., "Cracks at the Nanoscale," Int. J. Appl. Glass Sci. 4, 76 (2013).
31. Ciccotti M., "Stress-corrosion mechanisms in silicate glasses," J. Phys. D: Appl. Phys. 42 214006 (2009).
32. Mazer J. J. and Walther J. V., "Dissolution kinetics of silica glass as a function of pH between 40 and 85°C," J. Non-Cryst. Sol. 170, 32 (1994).
33. Fournier R. O., and Row J. J., "The solubility of amorphous silica in water at high temperatures and high pressures," Am. Mineralogist 62, 1052 (1977).

34. Dove P. M., Han N., Wallace A. F., and De Yoreo J. J., "Kinetics of amorphous silica dissolution and the paradox of the silica polymorphs," PNAS 105, 9903–9908 (2008).
35. Wakabayashi, H., Tomozawa, M., "Diffusion of Water into Silica Glass at Low Temperature," J. American Ceramic Society, 72 (10), 1850–1855 (1989).
36. Davis, K. M., Tomozawa, M., "Water Diffusion into Silica Glass: Structural Changes in Silica Glass and Their Effect on Water Solubility and Diffusivity," J. of Non-Crystalline Solids, 185 (3), 203–220 (1995).
37. Ge D., Yang L., Li Y., Zhao J., "Hydrophobic and thermal insulation properties of silica aerogel/epoxy composite," J. Non-Cryst. Solids 355, 2610–2615 (2009).
38. Sumetsky M., "Theory of SNAP devices: basic equations and comparison with the experiment," Opt. Express 20, 22537-22554 (2012).
39. Sumetsky M., "Optical bottle microresonators," Prog. Quant. Electron. 64, 1-30 (2019).
40. Hamidfar T., Tokmakov K. V., Mangan B. J., Windeler R. S., Dmitriev A. V., Vitullo D. L. P., Bianucci P., and Sumetsky M., "Localization of light in an optical microcapillary induced by a droplet," Optica 5, 382-388 (2018).
41. Bar-David D., Maayani S., Martin L. L., and Carmon T., "Cavity optofluidics: a μdroplet's whispering-gallery mode makes a μvortex," Opt. Express 26, 19115-19122 (2018).
42. Snyder A. and Love J., "Optical Waveguide Theory," Springer (1983).
43. Dmitriev A., Toropov N., and Sumetsky M., "Transient reconfigurable subangstrom-precise photonic circuits at the optical fiber surface," in 2015 IEEE Photonics Conf, IPC (2015).
44. Vitullo D. L. P., Zaki S., Gardosi G., Mangan B. J. Windeler R.S, Brodsky M., Sumetsky M., "Tunable SNAP microresonators via internal ohmic heating," Opt. Lett. 43, 4316 (2018).
45. Yarin L. P., Mosyak A., and Hetsroni G., "Fluid Flow, Heat Transfer and Boiling in Micro-Channels," Springer (2009).
46. Debenedetti P., "Metastable Liquids Concepts and Principles", Princeton Academic Press (1996).
47. Erné B. H. and Snetsinger P., "Thermodynamics of Water Superheated in the Microwave Oven," J. Chem. Ed. 77, 1309 (2000).
48. Karbalaei A., Kumar R. and Cho H. J., "Thermocapillarity in microfluidics—a review," Micromachines 7, 13 (2016).
49. Sumetsky M., "Delay of light in an optical bottle resonator with nanoscale radius variation: dispersionless, broadband, and low loss", Phys. Rev. Lett. 111, 163901 (2013).

50. Sumetsky M., "Management of slow light dispersion and delay time characteristics with SNAP bottle resonators," in CLEO 2014, OSA, paper STu3N.6 (2014),
51. Gorodetsky M. L., Pryamikov A. D., and Ilchenko V. S., "Rayleigh scattering in high-Q microspheres," J. Opt. Soc. Am. B 17, 1051 (2000).
52. Göstl R., Senf A. and Hecht S., "Remote-controlling chemical reactions by light: Towards chemistry with high spatio-temporal resolution," Chem. Soc. Rev. 43, 1982 (2014).
53. Houck, H.A., Du Prez, F.E., and Barner-Kowollik, C., "Controlling thermal reactivity with different colors of light," Nat. Commun. 8, 1869 (2017).
54. Sumetsky M., "Slow light optofluidics: a proposal," Opt. Lett. 39, 5578 (2014).

## SUPPLEMENTARY INFORMATION

### S1. Spectrogram treatment

The measured transmission power spectrograms feature a characteristic tilted quasi-periodic background shown in the exemplary Figure S1a. We suggest that this background originates from the interference of the fundamental and higher order modes propagating along the microfiber. The U-shaped microfibers used in our experiments had the waist diameter of approximately 1.7 μm and therefore are multimode waveguides. An optical mode propagating along the uniform microfiber with propagation constant $\beta_k$ has the form $E_k(x,\boldsymbol{r}) = A_k \exp(i\beta_k x) \cdot F_k(\boldsymbol{r})$ where $x$ is the coordinate along the microfiber and $\boldsymbol{r}$ is the two-dimensional vector in a microfiber cross-section[1]. In the MCF, this mode excites the WGM, which in the cylindrical coordinates has the form $\mathcal{E}_{k,m,p,s}(z,\rho,\phi) = C_k A_k \exp(i\beta_k x)\exp(im\phi) \cdot U_{m,p,s}(\rho) \cdot \Psi_{m,p,s}(z)$. Here $m$, $p$ and $s$ are the azimuthal, radial and polarization quantum numbers, $C_k$ are the coupling coefficients and constants $A_k$ are determined by the input optical power of the MF While the MF tapers used in our experiments are adiabatic, their U-shaped bends excite higher order modes. For example, if only the fundamental HE11 mode and the next TE01 mode with propagation constants $\beta_0$ and $\beta_1$ are propagating along MF1 then the WGM excited in the MCF is determined by their superposition as

$$\mathcal{E}(z,\rho,\phi) = [C_0 A_0 \exp(i\beta_0 x) + C_1 A_1 \exp(i\beta_1 x)] \exp(im\phi) \cdot U_{m,p,s}(\rho) \cdot \Psi_{m,p,s}(z). \qquad (1)$$

We estimate the difference of propagation constants $\Delta\beta = |\beta_0 - \beta_1| \sim 0.6\ \mu m^{-1}$ from Figure S2 showing the dependencies of propagation constants on the silica microfiber diameter for λ=1.55 μm. This estimate was obtained from a similar plot in [2] calculated for λ=0.633 μm by rescaling. We arrive at a characteristic period of oscillations along the waist length of MF1 $\Delta x = \frac{2\pi}{\Delta\beta} \sim 10$ μm following from Equation 1. This period should correspond to the background oscillation period in Figure S1a $\Delta z \sim 500$ μm. If we assume that MCF axis $z$ is slightly tilted from the direction $y$ of MF1 translation (see Figure 3), then the tilt angle $\theta = \frac{\Delta x}{\Delta z} \sim 0.02$ radians or 1 degree. Since the MCF is not perfectly straight, we suggest that this angle can change along the MCF length and, therefore, the period of background oscillations in Figure S1a may change along the MCF length as well.

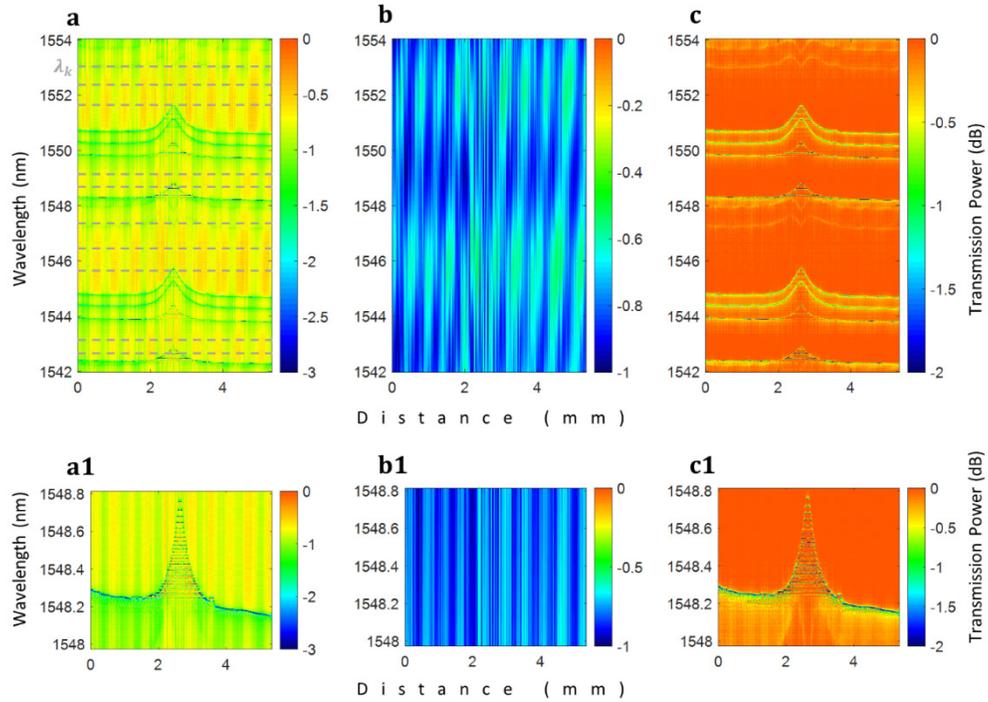

**Figure S1.** Spectrogram treatment. (a) The measured spectrogram with background. (b) The background, $P_{bkgr}(\lambda, z)$ calculated using spline interpolation over selected off resonant wavelengths for each z position. (c) The final spectrogram after subtraction of the background from the measured spectrogram. (a1-c1) Magnified parts of (a-c).

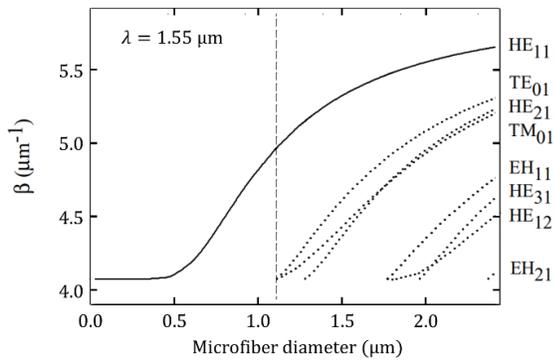

**Figure S2.** Propagation constants of different transmission modes of a silica microfiber as function of microfiber diameter at 1.55 μm wavelength (obtained by rescaling the results of [2]).

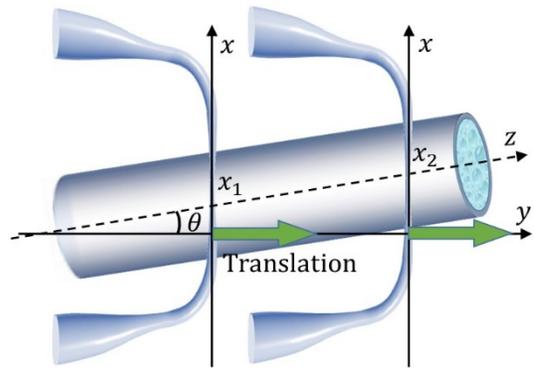

**Figure S3.** Illustration of a small angle between the MCF axis and MF1 translation direction, which gives rise to the quasi-periodic background observed.

We perform cleaning of the spectrogram shown in Figure S1a as follows. First, we choose 10 horizontal lines in this spectrogram corresponding to wavelengths, $k$=1,2...,10, which are situated in the areas where the propagation is not resonant i.e. away from the CWs. At these wavelengths, we expect the transmission power to be smooth. To this end, we find the off-resonant power and use spline interpolation at each z position. Then we generate the spectrogram, shown in Figure S2b. The corrected spectrogram shown in Figure S1c is the difference between the original and background spectrograms, $P_{corr}(\lambda, z) = P(\lambda, z) - P_{bkgr}(\lambda, z)$. Finally, Figures S1b1 and c1 show the magnified background and corrected spectrograms in the neighbourhood of the CW at 1548 nm.

## S2. Evidence of liquid water heating by evanescent WGMs

The slow cooking process described here is induced by the WGMs distributed along the MCF rather than by local heating of the optical fiber previously investigated in [3]. In order to justify this, we estimate the temperature variation along the fiber axis $z$ induced by local heating at MF2 contact position as $\Delta T(z) = \Delta T_0 \exp\left[-(2h/r_0)^{\frac{1}{2}}|z|\right]$, where $\Delta T_0$ is the variation of temperature at the heat source, $h$ is fiber-air heat transfer coefficient, and $k$ is the fiber heat conductivity. It can be shown that this temperature distribution fits the CW profile (proportional to $\Delta T(z)$) in Figure S4a measured in [3] for the silica optical fiber with $r_0$=19 μm, $k = 1.38\, W/m \cdot K$ and $h = 350\, W/m^2 \cdot K$. On the other hand, for a MCF with a thin wall considered here, the thermal conductivity is primarily determined by the water heat conductivity, while the value of $h$ should be greater than that determined for the bulk fiber due to the additional contribution of water. If we assume that the heat source is primarily localized near MF2 then the widest possible temperature distribution will be determined by the above equation for $\Delta T(z)$ where we use the water heat conductivity $k = 0.60\, W/m \cdot K, r_0 = 45$ μm, and the same $h$. Figure S4b compares this distribution (white curve) with the CW profile (proportional to the temperature distribution) we observed experimentally taken from Figure 2b1 of the main text. It is seen that the point source temperature distribution is much narrower than the CW profile measured for our MCF. Therefore, the observed temperature distribution is introduced by WGMs, which are spread along the MCF axis, rather than a localized heat source. To double check our conclusions, we performed heating of the empty MCF. The result shown in Figure S4c confirmed that the effect of heating without water is much weaker. In contrast to heating of the water-filled MCF, heating of the empty MCF using 100 mW, for the same period of time (12 hours), did not result in permanent alterations of the MCF.

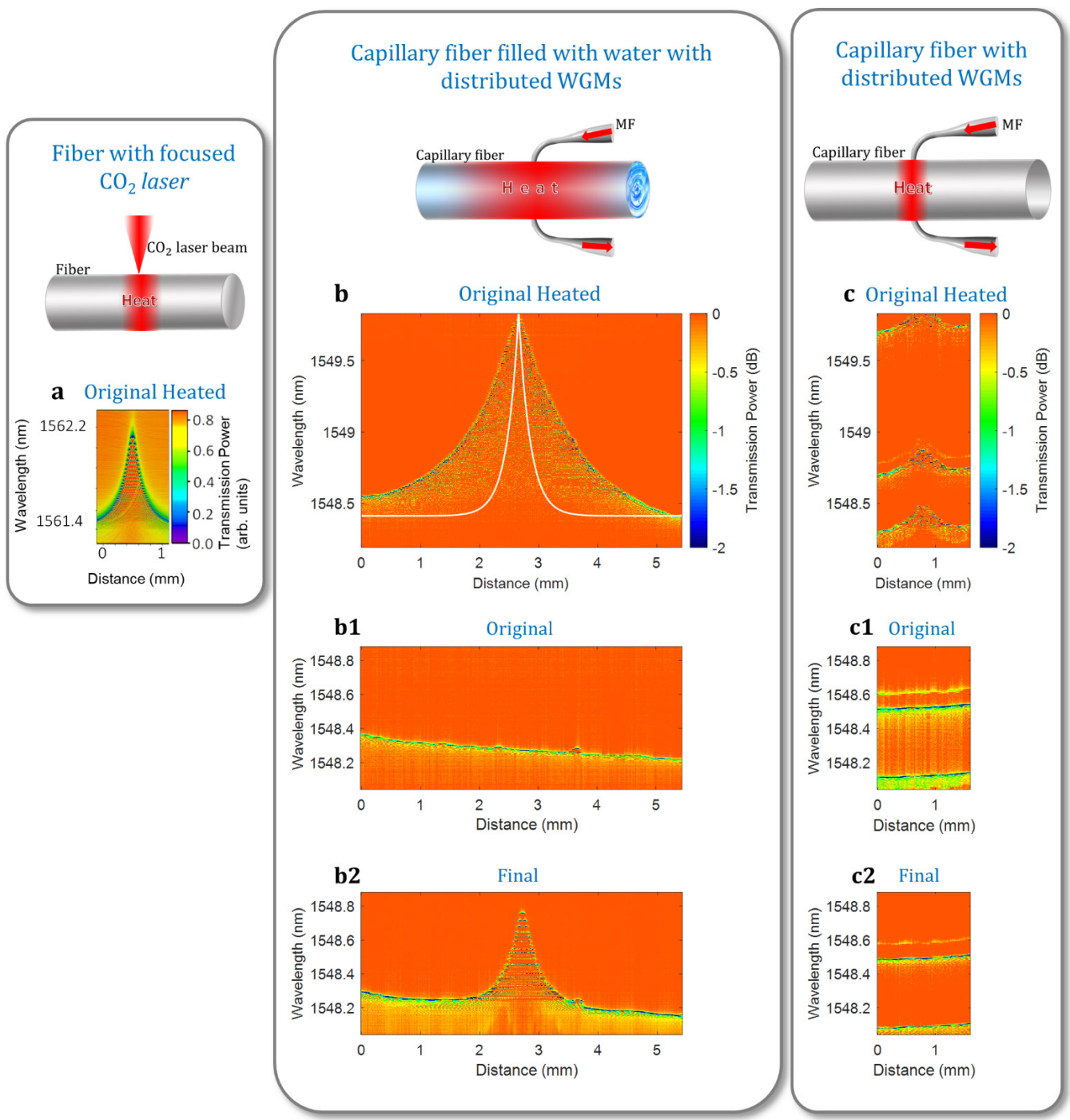

**Figure S4.** Local fiber heating vs. heating of a water-filled and empty MCF. (a) Spectrogram of a bulk fiber heated using a focused $CO_2$ laser beam[3]. (b) Spectrogram of the heated water-filled MCF from Figure 2 of the main text; (b1) original before heating; (b2) permanently induced microresonator after 12 hour heating. (c) Spectrogram of the heated empty MCF; (c1) original before heating; (c2) after 12 hour heating.

Additionally, we confirmed that, in the experiments described in the main text, water was maintained in the liquid rather than in the vaporized state. We recorded the temporal spectrogram of WGM transmission spectra and, simultaneously, the MCF images using a microscope camera. We observed that occasionally water evaporated, when the WGM heating power was large enough. The spectrogram in Figure S5a shows variation of the transmission power measured each minute at the same fixed axial coordinate of MF2. In this experiment, we observed the liquid-vapor transition characterized by significant shifts in CWs (see Figures S5a and S5b) after ~6.5hrs of heating. The images of the MCF before and after evaporation of water are shown in Figures 5c and 5d, respectively. We attribute the CW shifts to the change of the refractive index due to the disappearance of water as well as due to the related decrease in temperature described above (Figure 4b vs. Figure 4c). Since no similar CW shifts or microscope image changes were observed in the experiments of our paper, we confirm that water in these experiments was maintained in the liquid state.

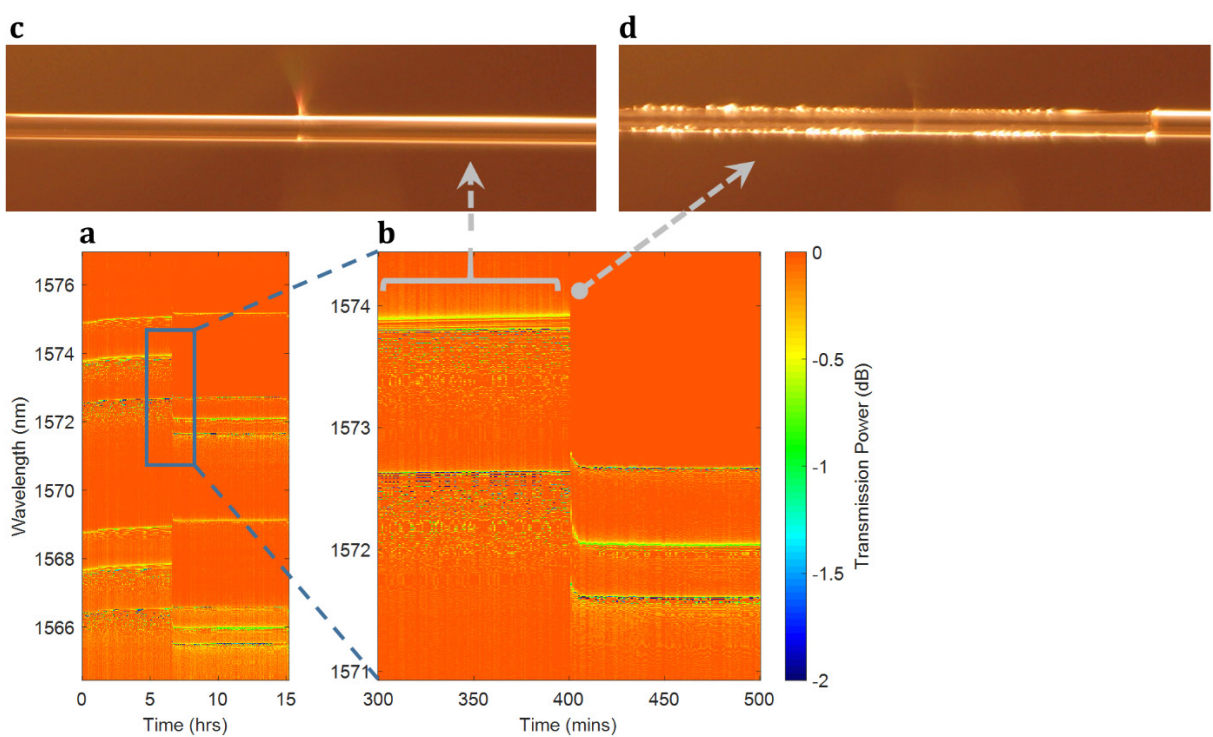

**Figure S5**. Evaporation of water. (a) The temporal spectrogram showing the variation of spectrum as a function of time at the position of MF2 in the process of slow cooking. (b) magnified part of (a) in the temporal region of evaporation. (c) Microscope image of water-filled MCF at $t < 6.5$ hrs. (d) Microscope image at ~6.5 hrs.

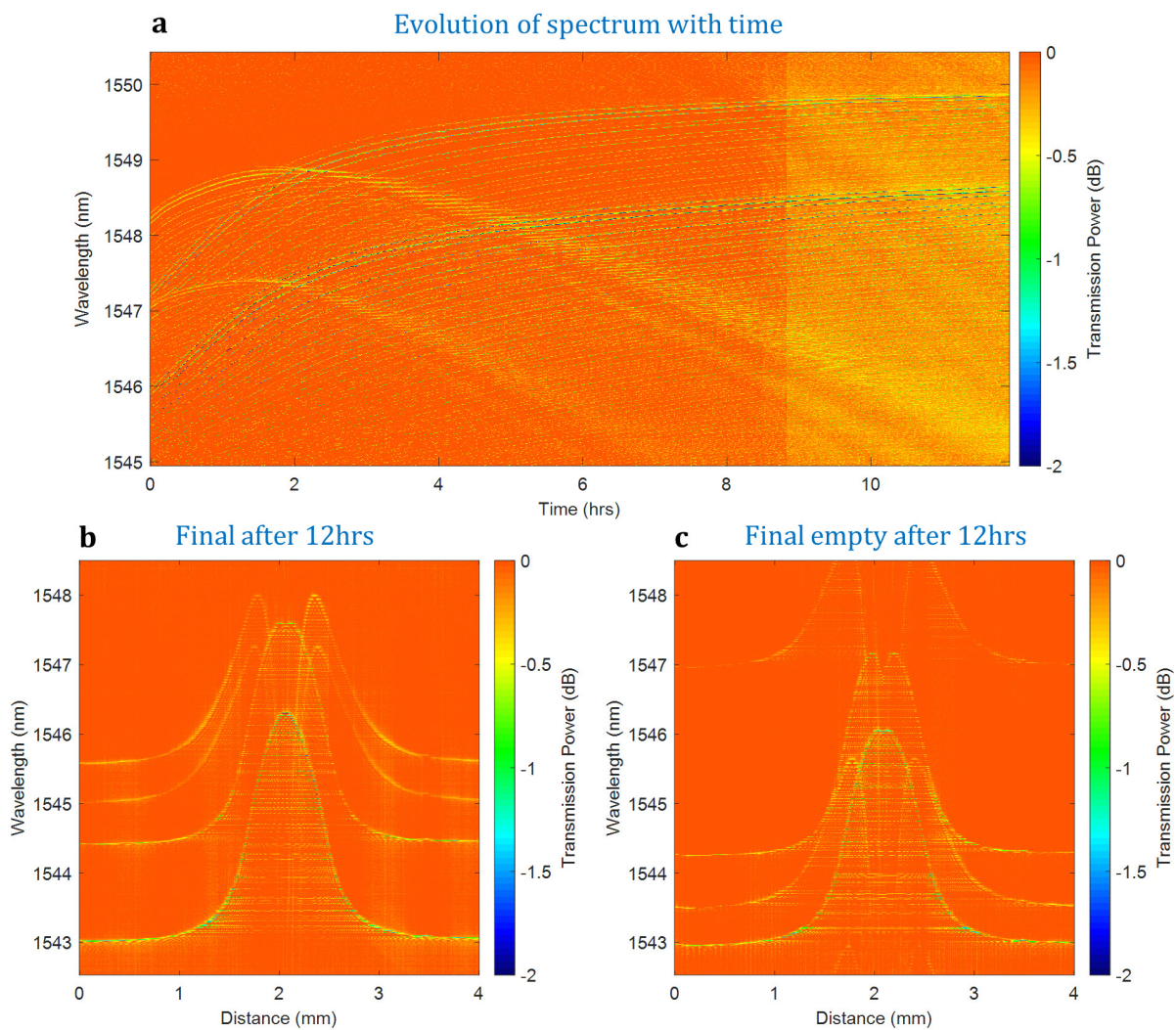

**Figure S6.** Some CWs experience non-monotonic evolution in time reaching the maximum, which results in permanently induced CWs having two maxima. (a) The temporal spectrogram showing the variation of CW as a function of time at the position of MF2 during the slow cooking process. (b) The spectrogram of the permanent alterations induced in the MCF filled with water. (c) The spectrogram of the permanent alteration after water is removed from the MCF and dried for 23 hours.

## S3. Additional experiments

The competition of different processes resulting in permanent alterations induced by the interaction between heated water and silica can be revealed from the two experiments shown in Figures S6 and S7. In order to understand the heating process better, for both experiments we increased the MF2 input heating power to 100 mW. The temporal spectrogram in Figure S6a, shows that initially the processes leading to the increase of CWs are dominant. Later after 2 hrs of heating,

CWs with smaller radial numbers continue to grow, while those with higher radial numbers achieve maximum and start decreasing (see CWs with less contrast in Figure S6a). After 12 hrs of heating, we measured the spectrograms of the unheated MCF with and then without water, shown in Figures S6b and S6c. The lower radial modes demonstrate negative CW variation near the MF2 contact position. This behavior demonstrates that at higher temperatures these CWs reach maximum earlier and then proceed to decrease.

In another experiment, shown in Figure S7, we measured the spectrograms of the same unheated MCF filled with water after 15 and then 37 hrs of slow cooking. In Figure S7a, we observe that the

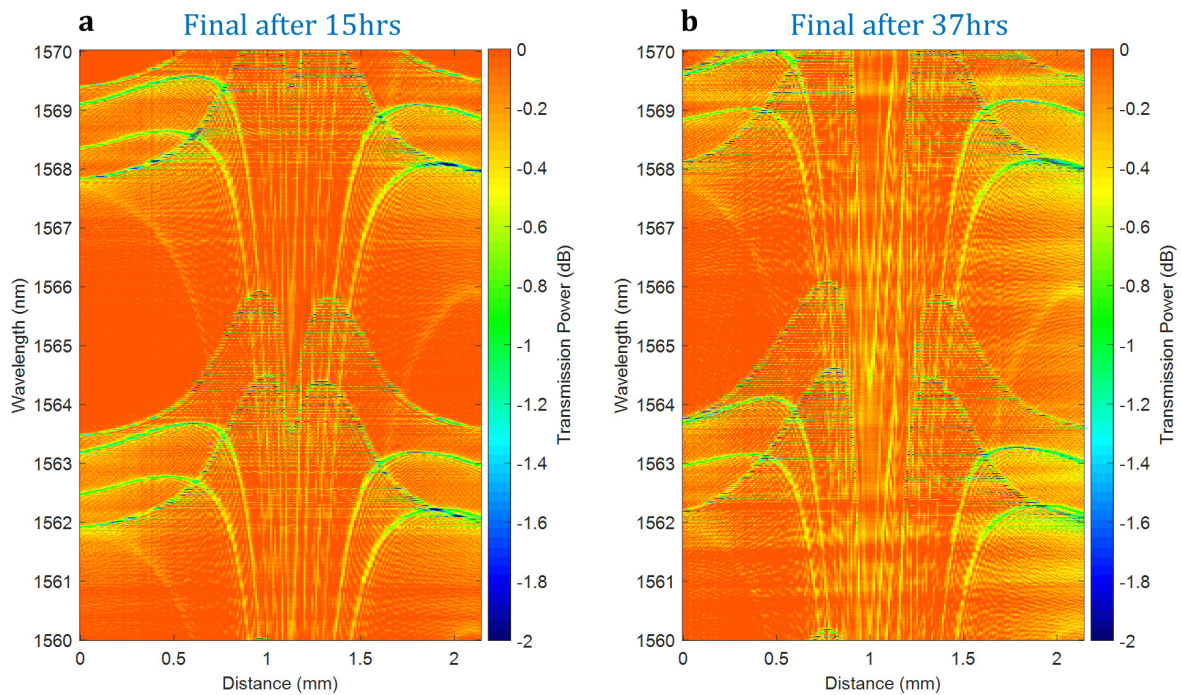

**Figure S7.** Longer heating at 100mW demonstrate that all CWs start to progress negatively. (a) The spectrogram of the water-filled MCF showing the permanent alterations after 15 hrs of slow cooking. (b) The spectrogram of the same water-filled MCF after further slow cooking for a total 37 hrs.

profiles of induced CW variations do not necessarily increase with the local temperature maintained during the slow cooking process. It is seen that WGMs with higher radial quantum numbers, which have a larger overlap with the region adjacent to the MCF internal surface[4], are more strongly altered than those with smaller radial numbers. Consequently, the regions of CWs with higher radial numbers in the vicinity of MF2 have larger negative variation growing with time. Comparing Figures

S7a and S7b, we see that longer heating both further decreases the CW, and extends this negative CW variation along the fiber away from the MF2 contact position.

Overall from Figures S6 and S7, given sufficient time and cooking temperature all WGMs observed are affected. Therefore the characteristic depth of the induced variation of refractive index, due to the propagation of the silica-water interaction processes into the silica wall, may achieve micrometer order. Further theoretical and experimental work is required to better understand these processes.

**REFERENCES**

1. Snyder A. and Love J., "Optical Waveguide Theory," Springer (1983).
2. Tong L., Lou J., and Mazur E., "Single-mode guiding properties of subwavelength-diameter silica and silicon wire waveguides," Opt. Express 12, 1025 (2004).
3. Dmitriev A., Toropov N., and Sumetsky M., "Transient reconfigurable subangstrom-precise photonic circuits at the optical fiber surface," in 2015 IEEE Photonics Conference (IPC), IEEE, pp. 1–2 (2015).
4. White I. M., Oveys H., and Fan X., "Liquid-core optical ring-resonator sensors," Opt. Lett. 31, 1319 (2006).